\documentclass[twocolumn]{article}

\usepackage{algorithm}
\usepackage{algpseudocode}
\usepackage{amsmath}
\usepackage{mathtools}
\usepackage{amssymb}

\usepackage{authblk}
\usepackage{url}
\usepackage{derivative}
\usepackage{caption}
\usepackage{subcaption}
\usepackage{hyperref}
\usepackage{float}
\usepackage{tikz}
\usetikzlibrary{arrows.meta, positioning, shapes.geometric}
\usepackage{enumitem}
\usetikzlibrary{positioning, arrows.meta}
\usepackage{multirow}
\usepackage{booktabs}
\usepackage{array}
\usepackage{bm}
\usepackage{fancyhdr}
\usepackage{xcolor}

\fancypagestyle{plain}{
  \fancyhf{} 

  \fancyfoot[C]{%
    \parbox{\textwidth}{\centering\color{gray}\small
    30$^\text{th}$ International Symposium on Space Flight Dynamics\\
    1--5 June 2026, Toulouse, France}
  }
}

\usepackage[explicit]{titlesec}
\renewcommand{\thesection}{\Roman{section}}

\titleformat{\section}{\large\scshape\centering}{\thesection.\space}{0pt}{#1}[]
\titlespacing*{\section}{0pt}{0.5\baselineskip}{0pt}
\titleformat{\subsection}{\normalsize\itshape}{\Alph{subsection}.\space}{0pt}{#1}[]
\titlespacing*{\subsection}{0pt}{0.5\baselineskip}{0pt}
\titleformat{\subsubsection}{\normalsize\itshape}{\arabic{subsubsection}.\space}{0pt}{#1}[]
\titlespacing*{\subsubsection}{0pt}{0.5\baselineskip}{0pt}
\usepackage{geometry}
\newgeometry{left=2cm, right=2cm, top=2.5cm,bottom=3.5cm}
\makeatletter
\renewcommand{\fnum@figure}{Fig. \thefigure}
\renewcommand{\fnum@table}{Tab. \thetable}
\makeatother

\title{\textbf{ Reachability for Low-Thrust Trajectories via Maximum Initial Mass}}
\author[(1)]{Giacomo Acciarini}
\author[(1)]{Dario Izzo}
\author[(2)]{Zhong Zhang}

\affil[(1)]{Advanced Concepts Team, European Space Agency, European Space Research and Technology Centre (ESTEC), Keplerlaan 1, 2201 AZ Noordwijk, The Netherlands}
\affil[(2)]{Politecnico di Milano, Piazza Leonardo da Vinci 32, Milan, Italy}
\affil[]{giacomo.acciarini@gmail.com}

\date{}  

\begin{document}
\maketitle

\begin{abstract}
\vspace{-1.3\baselineskip}
\textbf{\emph{\quad Abstract} - 
Reachability analysis plays a central role in low-thrust spacecraft trajectory optimization by identifying which target states can be achieved under constraints on time, thrust, and propellant. Traditional approaches construct reachable sets by solving numerous optimal control problems over grids of terminal states, requiring extensive forward simulations with fixed initial conditions. While effective, this method is computationally expensive and becomes impractical for high-dimensional systems or nonlinear dynamics, such as those encountered in cislunar environments or solar sail missions.
This work introduces a dual formulation of the reachability problem. Instead of directly computing the full reachable set, we determine, for fixed transfer time and boundary conditions, the maximum allowable initial mass (or, for solar sails, a scalar sail-strength parameter) that enables a successful transfer. A target is deemed reachable if the spacecraft’s initial mass does not exceed this threshold. This transforms the problem into a scalar optimization for each target, yielding a well-behaved scalar field that encodes the same feasibility information as the classical reachable set while exhibiting smoother numerical behavior.
We leverage indirect maximum-initial-mass (MIM) formulations for both electric low-thrust and solar-sail dynamics and show how they can be used as efficient reachability oracles. Building on this, we then construct data-driven surrogate models that approximate the MIM-based reachability indicator. In particular, we investigate fully connected architectures and show that residual networks provide the best trade-off between accuracy, training stability, and complexity in this setting. The resulting surrogates enable rapid reachability evaluation while preserving the advantageous numerical properties of the dual formulation, offering a practical tool for preliminary mission design and feasibility assessment.
}
\end{abstract}
\section{Introduction}
\label{sec:introduction}

Low-thrust propulsion technologies, such as electric thrusters and solar sails, enable highly efficient spacecraft trajectories at the price of complex, long-duration transfers governed by nonlinear dynamics and tight control constraints. In this context, \emph{reachability} questions, whether a given target state can be achieved within a prescribed time-of-flight, thrust budget, and propellant mass, are central to preliminary mission design, target selection, and trade-space exploration.

    Classical reachability analyses for low-thrust systems typically rely on solving many two-point boundary-value problems over grids of terminal states, thereby constructing approximations of the reachable set in the state space~\cite{holzinger2012reachability, pellegrino2021reachability, lee2018reachable, chen2021minimum}. These approaches require a large number of indirect or direct optimal-control solves and quickly become impractical for high-dimensional systems, long time horizons, or complex dynamical environments. Moreover, minimum-time formulations introduce strong nonlinearities and sensitivity near the reachable-set boundary, which can hinder convergence and complicate surrogate modeling. For this reason, there have been works that started proposing approximators of the reachable set~\cite{bowerfind2024rapid, wang2024analytical, lee2018reachable}.

In parallel, recent work has shown the usefulness of dual formulations based on maximum initial mass for low-thrust transfer design~\cite{hennes2016fast, mereta2017machine, acciariniComputingLowthrustTransfers2024a, izzo2025asteroid}, as well as the promise of machine-learning surrogates that replace repeated optimal-control solves by fast evaluations. However, an explicit treatment of maximum-initial-mass (MIM) reachability for general low-thrust systems, its properties, and its interplay with learned surrogates is still lacking.

The present work addresses this gap with two main contributions. First, we formulate reachability for low-thrust trajectories in terms of a maximum initial mass (or, for solar sails, a scalar sail-strength parameter) defined over fixed boundary conditions and time-of-flight. This yields an implicit scalar-field representation of the reachable set, which we analyze and compare to its "dual" minimum-time formulations, highlighting its improved numerical behavior and regularity. These maximum initial mass problems are then solved via indirect methods, leveraging Pontryagin Minimum Principle (PMP) for both low-thrust propulsion and solar-sail dynamics. Second, we construct and evaluate machine-learning surrogates for these dual problems, demonstrating the performances that residual networks with smooth activations (in particular residual networks with \texttt{Softplus} units) achieve for MIM-based reachability. Together, these elements provide an efficient and accurate framework for reachability analysis in low-thrust mission design.

\section{Reachability in Low-Thrust Trajectories}
\label{sec:reachability_in_low_thrust_trajectories}
The setting we study is cases in which the dynamics is driven by a state-dependent dynamics vector, $\pmb{f}$, and a control-dependent component, $\pmb{g}$:
\begin{equation}
    \dot{\pmb{x}} = \pmb{f}(\pmb{x}(t))+\pmb{g}(\pmb{x}(t),\pmb{u} )
\end{equation}
where $\pmb{x}\in\mathbb{R}^{n_x}$ and $\pmb{u}$ are the state and control vectors, respectively.
In cases where the mass of the spacecraft varies, we can find its time evolution augmenting the system with the following ordinary differential equation:
\begin{equation}
    \dot{m}= -\dfrac{T_{\textrm{max}}}{I_{sp}g_0}\|\pmb{u}(t)\|_2\text{,}
\end{equation}
where $\|\pmb{u}\|_2$ is the $l_2$-norm magnitude of the throttle vector, with $T_{\textrm{max}}$ being the maximum thrust allowable, $m_0$ is the initial mass, $g_0=9.80665$ m/s$^2$ the gravitational acceleration at the sea level of the Earth, and $I_{sp}$ the specific impulse. In our experiments, we will focus on two different scenarios: a low-thrust trajectory optimization case with electric propulsion, where the underlying dynamics is a two-body one, with added acceleration coming from the low-thrust engine, and a solar sailing case, in which there is no mass loss due to propellant, as the thrust is generated leveraging the solar radiation pressure of the Sun. Both these cases will be detailed in Sec.~\ref{sec:electric_low_thrust_propulsion} and \ref{sec:solar_sailing}.

Classically, the reachable set, $\mathcal{R}$, is then defined as the collection of states that can be reached from an initial state and time, with a fixed duration $[t_0, t_f]$, with a control that is admissible (i.e., that satisfies the constraints) $\pmb{u}\in \mathcal{U}$:
\begin{equation}
\begin{aligned}
\mathcal{R}(t_f;\pmb{x}_0,m_0)
:=
\Big\{ &
\pmb{x}_f \in \mathbb{R}^{n_x}
\ \Big|\ 
\exists\, \pmb{u}(\cdot)\in\mathcal{U}
\text{ s.t.} \\
& \dot{\pmb{x}}(t)=\pmb{f}(\pmb{x}(t))
+\pmb{g}(\pmb{x}(t),\pmb{u}(t)), \\
& \dot{m}(t)=
-\dfrac{T_{\max}}{I_{sp}g_0}\|\pmb{u}(t)\|_2, \\
& \pmb{x}(t_0)=\pmb{x}_0,\quad m(t_0)=m_0, \\
& \pmb{x}(t_f)=\pmb{x}_f
\Big\}.
\end{aligned}
\end{equation}

Usually, solving this problem is computationally intensive, as it requires exploring a high-dimensional space of admissible controls and terminal states. In practice, this often involves discretizing the target state space and solving a large number of boundary-value or optimal control problems, each subject to nonlinear dynamics, control bounds, and, potentially, mass depletion effects. Moreover, the structure of $\mathcal{R}$ can be highly nonconvex and sensitive to the choice of final time, making its accurate characterization, especially at the boundaries, very challenging, as discussed profusely in the literature~\cite{holzinger2012reachability, pellegrino2021reachability, lee2018reachable, chen2021minimum}. In fact, the classical reachability formulation reduces the problem to a binary feasibility one, with a hard discontinuous boundary: causing numerical difficulties particularly for the solutions of the two point boundary value problem that lay at the boundary of the set. These limitations encourage the introduction of alternative formulations that shift the focus from explicitly constructing the reachable set to evaluating reachability through scalar quantities, which are possibly defined everywhere and continuous (and differentiable).

\paragraph{Dual formulation via maximum initial mass.}
Motivated by this, we formulate the problem as a maximum initial mass one, where the objective is, given an initial and final state ($\pmb{x}_0$ and $\pmb{x}_f$, respectively) and a transfer window time $[t_0, t_f]$, to find the maximum initial mass that makes the transfer possible:
\begin{equation}
\begin{aligned}
m_{\textrm{MIM}}(\pmb{x}_0,\pmb{x}_f,t_f)
:= \sup \Big\{& m_0 \in \mathbb{R} \ \Big|\ 
 \exists\, \pmb{u}(\cdot)\in\mathcal{U} \text{ s.t. } \\
& \dot{\pmb{x}}(t)=\pmb{f}(\pmb{x}(t))
+\pmb{g}(\pmb{x}(t),\pmb{u}(t)), \\
& \dot{m}(t)=
-\dfrac{T_{\max}}{I_{sp}g_0}\|\pmb{u}(t)\|_2, \\
& \pmb{x}(t_0)=\pmb{x}_0,\ m(t_0)=m_0, \\
& \pmb{x}(t_f)=\pmb{x}_f
\Big\}\text{,}
\end{aligned}
\end{equation}
where $m_f$ is dictated by the control history and mass-flow law. So instead of fixing $m_0$ and asking ``is $\pmb{x}_f$ reachable?'', we fix $\pmb{x}_0, \pmb{x}_f$, $t_f$ and ask ``what is the largest $m_0$ that still allows reaching $\pmb{x}_f$?''. Under \emph{mild} assumptions on $\pmb{g}$, the latter problem admits a solution, albeit with mass values that may not be feasible with the given spacecraft mass. Note that while this reduces the reachability problem to a threshold condition on the mass, the two problems are formally equivalent, and the function $m_{\textrm{MIM}}$ encodes the same feasibility information as the boundary of the reachable set.
This provides an implicit scalar-field representation of the reachable set, more analogous to value-function representations in Hamilton–Jacobi analysis~\cite{kirk2004optimal, bansal2017hamilton, fisac2015reach}. Level sets of this scalar field correspond to constant mass margins, and the reachable set boundary for a spacecraft of mass $m_{\mathrm{sc}}$ is obtained by thresholding at $m_{\textrm{MIM}}=m_{\mathrm{sc}}$.
\paragraph{Relation to minimum-time reachability.}
An important observation is that minimum‑time solutions also characterize the boundary of the reachable set~\cite{taheri2020many}. In the classical minimum‑time setting one fixes $(\mathbf{x}_0,\mathbf{x}_f,m_0)$ and minimizes the final time $t_f$, obtaining a time‑to‑reach value function on the target space. This formulation is often numerically more challenging than the maximum‑initial‑mass approach, due to the high sensitivity of the target state with respect to the optimal transfer time: small changes in the target state can induce large changes in the optimal time‑of‑flight, especially near the reachable‑set boundary. As a result, the time‑to‑reach function can exhibit sharper gradients and stronger nonsmoothness, and the associated two‑point boundary‑value problem (TPBVP) is often more prone to convergence issues and ill‑conditioning.
In contrast, in the present MIM formulation the time horizon is fixed and the unknowns are the initial mass and the costates, rather than the final time and the costates: the shooting function remains of the same dimension but is typically better conditioned because the time‑scaling freedom is removed. The control structure remains similar to the time‑optimal case, but the scalar quantity of interest is $m_{\textrm{MIM}}$ rather than $t_f$. Empirically, this dual scalar field varies more smoothly with the target state in practical low‑thrust problems~\cite{hennes2016fast, mereta2017machine, acciariniComputingLowthrustTransfers2024a, izzo2025asteroid, zhong2026}, which makes it better suited as a regression target and as a basis for surrogate reachability models.
\section{The Maximum Initial Mass Problem}

\subsection{Electric Low-Thrust Propulsion}
\label{sec:electric_low_thrust_propulsion}

We consider the controlled dynamics introduced in Sec.~\ref{sec:reachability_in_low_thrust_trajectories}, where the drift and control-dependent terms are denoted by $\pmb{f}$ and $\pmb{g}$, and the spacecraft mass evolves according to the low-thrust mass-flow law. The state is augmented with the mass,
\begin{equation}
    \pmb{z}(t)
    =
    \begin{bmatrix}
        \pmb{x}(t)\\
        m(t)
    \end{bmatrix}
    \in \mathbb{R}^{n_x+1},
    \label{eq:aug_dynamics_lt}
\end{equation}
and the dynamics reads
\begin{equation}
\begin{aligned}
\dot{\pmb{r}}(t)&=\pmb{v}(t)\\
    \dot{\pmb{v}}(t) &= -\mu \dfrac{\pmb{r}(t)}{r^3(t)}
+ \dfrac{T_{\textrm{max}}}{m}\pmb{u}(t),\\
    \dot{m}(t) &= -\dfrac{T_{\max}}{I_{sp}g_0}\,\lVert\pmb{u}(t)\rVert_2,
\end{aligned}
\end{equation}
where $\pmb{a}_g(\pmb{x})$ denotes the gravitational acceleration (two-body in this work) and we used the standard low-thrust relation $\pmb{g}(\pmb{x},\pmb{u}) = \tfrac{T_{\max}}{m}\,\pmb{u}$, with admissible controls $\pmb{u}(\cdot)\in\mathcal{U}$ satisfying the thrust magnitude bound $\lVert\pmb{u}(t)\rVert_2\le 1$.

For a fixed transfer time $[t_0,t_f]$ and fixed boundary states $\pmb{x}(t_0)=\pmb{x}_0$, $\pmb{x}(t_f)=\pmb{x}_f$, the maximum initial mass problem is
\begin{equation}
\begin{aligned}
\max_{m_0,\;\pmb{u}(\cdot)}\quad & m_0\\[0.2em]
\text{s.t.}\quad
& \dot{\pmb{r}}(t) = \pmb{v}(t),\\
& \dot{\pmb{v}}(t) = -\mu\dfrac{\pmb{r}(t)}{r^3(t)}
+ \dfrac{T_{\textrm{max}}}{m}\pmb{u}(t),\\
& \dot{m}(t) = -\dfrac{T_{\max}}{I_{sp}g_0}\,\lVert\pmb{u}(t)\rVert_2,\\
& \pmb{x}(t_0) = \pmb{x}_0,\quad \pmb{x}(t_f)=\pmb{x}_f,\\
& m(t_0) = m_0,\\
& \lVert\pmb{u}(t)\rVert_2 \le 1 \quad \forall t\in[t_0,t_f].
\end{aligned}
\label{eq:ocp_mim_lt}
\end{equation}
Reachability is then encoded by the scalar field $m_{\textrm{MIM}}(\pmb{x}_0,\pmb{x}_f,t_f)$: a target state $\pmb{x}_f$ is reachable for a given spacecraft iff its actual initial mass does not exceed this maximum admissible initial mass.

\subsubsection{Pontryagin minimum principle conditions}

We recast~\eqref{eq:ocp_mim_lt} as a minimization problem by considering the Mayer cost $\Phi$
\begin{equation}
    J = \Phi = -m(t_0) = -m_0,
\end{equation}
and apply Pontryagin’s Minimum Principle (PMP) in its standard form for fixed-time, endpoint-constrained problems~\cite{pontryagin1962mathematical}.
We split the state as $\pmb{x} = [\pmb{r}^{\mathsf T},\pmb{v}^{\mathsf T}]^{\mathsf T}$, with position $\pmb{r}$ and velocity $\pmb{v}$, and introduce costates $\pmb{\lambda}_r(t)$, $\pmb{\lambda}_v(t)$, and $\lambda_m(t)$ associated with $\pmb{r}(t)$, $\pmb{v}(t)$, and $m(t)$, respectively.
The Hamiltonian reads
\begin{equation}
\begin{aligned}
H\bigl(\pmb{x},m,\pmb{\lambda}_r,\pmb{\lambda}_v,\lambda_m,\pmb{u}\bigr)
&=\pmb{\lambda}_r^{\mathsf T}\pmb{v}
+ \pmb{\lambda}_v^{\mathsf T}
\bigl(\pmb{a}_g(\pmb{r})
+ \tfrac{T_{\max}}{m}\,\pmb{u}\bigr)\\
& \quad - \lambda_m \dfrac{T_{\max}}{I_{sp}g_0}\,\|\pmb{u}\|_2,
\end{aligned}
\label{eq:H_low_thrust}
\end{equation}
where $\pmb{a}_g(\pmb{r})$ denotes the gravitational acceleration (two-body in this work).

The transversality conditions follow from the fact that $J$ depends only on the initial mass, while the initial and final states $\pmb{x}(t_0)$ and $\pmb{x}(t_f)$ are fixed.
We obtain
\begin{equation}
    \lambda_m(t_0) = -\frac{\partial \Phi}{\partial m_0} = 1,
\end{equation}
while $\lambda_m(t_f) = 0 $.
All components of $\pmb{\lambda}_r$ and $\pmb{\lambda}_v$ at both endpoints are free and are determined by enforcing the two-point boundary conditions.

The remaining costate dynamics are given, as usual, by the negative partial derivatives of $H$ with respect to the state variables.
Because $\pmb{a}_g$ does not depend on $m$, the costate associated with the mass has the explicit evolution law
\begin{equation}
    \dot{\lambda}_m(t)
    =
    \pmb{\lambda}_v^{\mathsf T}(t)\,\frac{T_{\max}}{m^2(t)}\,\pmb{u}(t).
    \label{eq:lambda_m_dot}
\end{equation}
As we discuss next, the optimal control direction aligns with $-\pmb{\lambda}_v$, so $\pmb{\lambda}_v^{\mathsf T}\pmb{u}\le 0$ along optimal trajectories.
Since $T_{\max}>0$ and $m(t)>0$, this implies that $\dot{\lambda}_m(t)\le 0$ almost everywhere, i.e.\ the mass costate is monotonically decreasing and starts from $\lambda_m(t_0)=1$.
\subsubsection{Optimal control structure}

The Hamiltonian~\eqref{eq:H_low_thrust} is affine in the control $\pmb{u}$, subject to the thrust magnitude bound $\|\pmb{u}(t)\|_2\le 1$.
Minimization of $H$ yields an optimal thrust direction aligned with the velocity costate,
\begin{equation}
    \hat{\pmb{u}}^*(t)
    =
    -\frac{\pmb{\lambda}_v(t)}
    {\|\pmb{\lambda}_v(t)\|_2},
    \qquad
    \|\pmb{u}^*(t)\|_2 = 1,
\end{equation}
whenever $\pmb{\lambda}_v(t)\neq \pmb{0}$; see previous work for detailed derivations of the optimal thrust structure in thrust-bounded low-thrust transfers~\cite{kirk2004optimal, perez2018fuel, sidhoum2024indirect} .

Introducing the usual switching function $\sigma(t)$ that multiplies the thrust magnitude, one can show that, for the MIM formulation considered here and for the boundary conditions of interest, $\sigma(t)$ remains strictly negative along optimal trajectories.
Consequently, the optimal solution uses maximum available thrust everywhere.

It should be noted that, under the full-thrust structure together with a fixed transfer time, the propellant consumption can be directly determined as \( \Delta m = m_0 - m_f = (t_f - t_0){T_{\max}}/{I_{sp} g_0} \). Therefore, an alternative shooting setting can be adopted: since \( m_f \) is directly determined by \( m_0 \), the terminal mass is fixed, such that the unknown variable is effectively changed from \( m_f \) to \( \lambda_{m_f} \). This alternative formulation yields the same optimal trajectory, although \( \lambda_m \) may differ by a scaling factor. The equivalence between these two settings arises from the full-thrust structure, which is commonly derived in time-optimal control problems. A detailed explanation is beyond the scope of the present paper; some relevant discussions can be found in~\cite{guo2023}, while a more comprehensive analysis will be presented in future work.

\subsubsection{Forward–backward shooting}

In the numerical implementation, the PMP boundary-value problem is solved in a forward–backward fashion using a midpoint-matching shooting formulation~\cite{sidhoum2024indirect, jiang2012practical, perez2018fuel} and leveraging the smoothing techniques for the control and shooting functions discussed in ~\cite{bertrand2002new}. The state $\pmb{x}$ is represented in modified equinoctial elements, and the dynamics and adjoint equations are integrated up to the mismatch point, both in forward time from $t_0$ and in backward time from $t_f$. The forward-backward shooting method has proven very effective both for direct methods~\cite{izzo2026practical} and for indirect ones~\cite{sidhoum2024indirect, zhong2026_issfd}.

Let $\pmb{\lambda}_{\pmb{r},0}$, $\pmb{\lambda}_{\pmb{v},0}$ denote the initial position and velocity costates at $t_0$, and $\pmb{\lambda}_{\pmb{r},f}$, $\pmb{\lambda}_{\pmb{v},f}$ the corresponding costates at $t_f$.
The decision vector collects the unknown initial mass and costates at both boundaries:
\begin{equation}
    \pmb{X}
    =
    \bigl[
        \pmb{\lambda}_{\pmb{r},0}^{\mathsf T},
        \pmb{\lambda}_{\pmb{v},0}^{\mathsf T},
        m_0,
        \pmb{\lambda}_{\pmb{r},f}^{\mathsf T},
        \pmb{\lambda}_{\pmb{v},f}^{\mathsf T},
        m_f
    \bigr]^{\mathsf T}.
\end{equation}
For a given $\pmb{X}$, the augmented state and costates are integrated from $t_0$ to the midpoint $t_m = \tfrac{1}{2}(t_0+t_f)$ and from $t_f$ backward to $t_m$.
Continuity of both state and costate at $t_m$ then defines a system of nonlinear equations
\begin{equation}
    \pmb{F}(\pmb{X})
    =
    \pmb{z}^{\text{fwd}}(t_m;\pmb{X})
    - \pmb{z}^{\text{bck}}(t_m;\pmb{X})
    = \pmb{0},
\end{equation}
where $\pmb{z} = [\pmb{x}^{\mathsf T},m,\pmb{\lambda}_r^{\mathsf T},\pmb{\lambda}_v^{\mathsf T},\lambda_m]^{\mathsf T}$ collects state and costates.
Solving $\pmb{F}(\pmb{X})=\pmb{0}$ provides the costate boundary conditions and the associated maximum initial mass $m_{\textrm{MIM}}$ for the prescribed transfer.
In the numerical solver, this root-finding problem is tackled using a variant of the Minpack hybrd algorithm~\cite{powell1970hybrid}.
\subsection{Solar Sailing}
\label{sec:solar_sailing}

For the solar-sail case we consider a spacecraft subject to the gravitational attraction of the central body and to a radiation-pressure acceleration generated by an ideal flat sail.
The state is again
\begin{equation}
    \pmb{x}(t)
    =
    \begin{bmatrix}
        \pmb{r}(t)\\
        \pmb{v}(t)
    \end{bmatrix}
    \in\mathbb{R}^6,
\end{equation}
with $\pmb{r}$ and $\pmb{v}$ the position and velocity in an inertial frame.
The control is the sail normal direction $\pmb{i}(t)\in\mathbb{R}^3$, constrained by
\begin{equation}
    \lVert\pmb{i}(t)\rVert_2 = 1,
\end{equation}
and the sail strength is parametrised by a scalar coefficient $k_0>0$ proportional to the sail lightness number.

Let $r(t)=\lVert\pmb{r}(t)\rVert_2$ and $\pmb{u}_r(t)=-\pmb{r}(t)/r(t)$ denote the radial unit vector pointing towards the Sun.
The sail acceleration model~\cite{mcinnes2004solar} adopted in this work is
\begin{equation}
\begin{aligned}
    \dot{\pmb{r}}(t) &= \pmb{v}(t),\\
    \dot{\pmb{v}}(t) &= -\mu\,\frac{\pmb{r}(t)}{r^3(t)}
    - k_0\left(\frac{r_0}{r(t)}\right)^2 \cos^2\alpha(t)\,\pmb{i}(t),
\end{aligned}
\end{equation}
where $\mu$ is the standard gravitational parameter, $r_0$ is a reference distance, and $\alpha(t)$ is the cone angle between $\pmb{i}(t)$ and $\pmb{u}_r(t)$, so that $\cos\alpha(t)=\pmb{i}^{\mathsf T}(t)\pmb{u}_r(t)$.
The control appears only through the sail orientation $\pmb{i}(t)$, while $k_0$ is treated as a scalar design parameter scaling the solar-radiation-pressure acceleration (and inversely proportional to the sail mass).

We consider a fixed time of flight $[t_0,t_f]$, with the initial state $(\pmb{r}_0,\pmb{v}_0)$ prescribed and the final position constrained to a given target $\pmb{r}_f$, while the terminal velocity is free.
To obtain a scalar reachability descriptor analogous to the maximum initial mass in the low-thrust case, we use the sail-strength parameter $k_0$ itself as the optimisation variable and define a Lagrange cost as:
\begin{equation}
    J 
    =
    \int_{t_0}^{t_f} k_0\,\mathrm{d}t,
\end{equation}
so that we minimise $k_0$ for a fixed transfer time (equivalently, we maximise the sail mass for a given sail area).
In the following, we augment the state by treating $k_0$ as a constant state satisfying $\dot k_0=0$.

\subsubsection{Pontryagin minimum principle}

We introduce costates $\pmb{\lambda}_r(t)$ and $\pmb{\lambda}_v(t)$ associated with $\pmb{r}(t)$ and $\pmb{v}(t)$, respectively, and augment the constant state $k_0$, and introduce the associated auxiliary costate $I_{k_0}(t)$ to encode the optimality condition with respect to $k_0$.
The augmented state--costate is
\begin{equation}
    \pmb{z}(t)
    =
    \begin{bmatrix}
        \pmb{r}(t)\\
        \pmb{v}(t)\\
        k_0\\
        \pmb{\lambda}_r(t)\\
        \pmb{\lambda}_v(t)\\
        I_{k_0}(t)
    \end{bmatrix}.
\end{equation}
The Hamiltonian for the minimisation problem can be expressed as
\begin{equation}
\begin{aligned}
H
&=
\pmb{\lambda}_r^{\mathsf T}\pmb{v}
+ \pmb{\lambda}_v^{\mathsf T}
\left(
    -\mu\,\frac{\pmb{r}}{r^3}
    - k_0\left(\frac{r_0}{r}\right)^2\cos^2\alpha\,\pmb{i}
\right)
+ k_0,
\end{aligned}
\label{eq:H_solar_sail}
\end{equation}
with $\cos\alpha = \pmb{i}^{\mathsf T}\pmb{u}_r$. Since $\dot k_0=0$, the term $I_{k_0}\dot k_0$ vanishes identically and does not appear explicitly in the Hamiltonian.

The additional state and costate equations associated with $k_0$ are 
$\dot k_0 = 0$,  and  $\dot I_{k_0} = -{\partial H}/{\partial k_0}$. 
 From \eqref{eq:H_solar_sail}, we obtain 
 \begin{equation} 
 \dot I_{k_0} = -1 + \left(\frac{r_0}{r}\right)^2 \cos^2\alpha\, \pmb{\lambda}_v^{\mathsf T}\pmb i . 
 \end{equation} 
 Since $k_0$ is unknown and free, the associated transversality conditions are \begin{equation} I_{k_0}(t_0)=0, \qquad I_{k_0}(t_f)=0. \end{equation} These two boundary conditions enforce the first-order optimality of the constant parameter $k_0$.

The other costate equations are obtained as usual from the partial derivatives of the Hamiltonian w.r.t the state and are integrated together with the state and $I_{k_0}(t)$.

\subsubsection{Optimal sail orientation}

The control enters the Hamiltonian only through the sail term,
\begin{equation}
    H_i
    =
    - k_0\left(\frac{r_0}{r}\right)^2
    \cos^2\alpha\,\pmb{\lambda}_v^{\mathsf T}\pmb{i},
\end{equation}
subject to the constraint $\lVert\pmb{i}\rVert_2=1$.
Minimising $H$ with respect to $\pmb{i}$ yields a closed-form expression for the optimal sail normal $\pmb{i}^*$ as a function of the costates and the radial direction~\cite{mcinnes2004solar}.
In particular, the optimal \emph{cone angle} $\alpha^*$ and \emph{clock angle} define a unit vector $\pmb{i}^*$ in the plane spanned by the radial unit vector $\pmb{u}_r$ and the transverse component of $\pmb{\lambda}_v$.
The resulting law coincides with the classical optimal orientation for ideal solar sails, whereby the sail normal lies in the plane generated by $\pmb{u}_r$ and $\pmb{\lambda}_v$ and balances the radial and transverse contributions to minimise the Hamiltonian~\cite{mcinnes1999solar}.

\subsubsection{Single shooting}

As in the electric low‑thrust case, the necessary conditions for optimality are enforced via a shooting formulation.
In this case, however, the control structure is smooth, and the associated costate dynamics exhibit reduced sensitivity to the initial costate values.
As a result, we do not find forward–backward schemes strictly necessary to regularize the boundary‑value problem, for the time horizons and transfer types that we investigate.
In our numerical experiments, the single‑shooting formulation converged reliably, which motivated its adoption in the solar‑sail case.

The decision vector collects the initial costates and the parameter $k_0$,
\begin{equation}
    \pmb{X}
    =
    \bigl[
        \pmb{\lambda}_r^{\mathsf T}(t_0),
        \pmb{\lambda}_v^{\mathsf T}(t_0),
        k_0
    \bigr]^{\mathsf T},
\end{equation}
and the augmented PMP dynamics is integrated over the fixed time interval $[t_0,t_f]$.
The residuals enforce terminal position and velocity matching and the parameter-stationarity condition,
\begin{equation}
\begin{aligned}
    \pmb{r}(t_f;\pmb{X}) - \pmb{r}_f &= \pmb{0},\\
    \pmb{v}(t_f;\pmb{X}) - \pmb{v}_f &= \pmb{0},\\
     I_{k_0}(t_f;\pmb{X}) &= 0,
\end{aligned}
\end{equation}
resulting in a system of seven nonlinear equations in seven unknowns.
As before, this root-finding problem is solved via the Minpack hybrd algorithm, yielding the optimal parameter $k_0$ and the associated sail trajectory and costate history.
The corresponding maximum mass (or minimum sail area) for the chosen dynamical model can then be obtained by inverting the relationship between $k_0$ and the characteristic sail acceleration.
\section{Machine Learning Surrogates}
\label{sec:ml_surrogates}
We now construct data–driven surrogate models for the maximum–initial–mass formulation introduced above.
The overarching goal is to replace repeated indirect optimal–control solves by fast evaluations, both for the two–body low–thrust reachability problem and for the solar–sail transfers, in the spirit of previous work on neural approximators for asteroid–belt transfers~\cite{hennes2016fast,mereta2017machine,acciariniComputingLowthrustTransfers2024a, zhong2026}.
A key contribution of this paper is to show that residual multilayer perceptrons, equipped with smooth activations, provide particularly effective surrogates for these dual reachability indicators, combining high accuracy with relatively modest parameter counts.

\subsection{Two–body low–thrust  reachability surrogate}
\label{sec:ml_surrogate_2bp}

In the two–body low–thrust setting we do not regress a continuous performance index, but directly learn a binary reachability indicator.

A transfer is labelled reachable if the maximum initial mass exceeds a prescribed spacecraft mass \(m_{\mathrm{sc}}\) (here \(m_{\mathrm{sc}}=1500\ \mathrm{kg}\)), and unreachable otherwise.
Formally, the classification label is
\begin{equation}
    y
    =
    \mathbb{I}\!\left( \frac{m_{\text{MIM}}}{m_{\mathrm{sc}}} \ge 1 \right)
    \in \{0,1\},
\end{equation}
so that \(y=1\) denotes a reachable target, whereas \(y=0\) corresponds to an infeasible transfer.

For each transfer, we solve a forward-backward shooting indirect formulations, described in Sec.~\ref{sec:electric_low_thrust_propulsion}, and a set of normalised features, and the true maximum initial masses are then stored.

The input feature vector \(\bm{\phi}\in\mathbb{R}^{11}\) concatenates
\begin{itemize}
    \item the normalised time of flight and specific impulse, denoted \(\texttt{tof\_n}\) and \(\texttt{Isp\_n}\);
    \item the three components of the Lambert initial \(\Delta\bm{v}\) at the departure asteroid, expressed in the local RTN basis and normalised (\texttt{dv\_lambert\_start\_r\_n}, \texttt{\_t\_n}, \texttt{\_n\_n});
    \item the three components of the Lambert arrival \(\Delta\bm{v}\) at the target asteroid in the same RTN frame (\texttt{dv\_lambert\_target\_r\_n}, \texttt{\_t\_n}, \texttt{\_n\_n});
    \item the normalised eccentricity of the departing orbit (\texttt{ecc\_n});
    \item the cosine and sine of the true anomaly at departure (\texttt{cos\_f}, \texttt{sin\_f}).
\end{itemize}
All features are standardised based on the dataset statistics and subsequently clipped to a finite range, which improves conditioning and limits the influence of outliers. As shown in previous works, rotational invariant and Lambert solutions inputs are very useful to improve the accuracy of the surrogate~\cite{acciariniComputingLowthrustTransfers2024a}. On the other hand, the feature space is also maintained as compact as possible, as this has been also shown to have benefits~\cite{zhong2026}.

We assemble the dataset into disjoint training, validation, and test sets with proportions \(80\%\), \(10\%\), and \(10\%\), respectively.
The class label distribution is not imbalanced, as roughly half of the random asteroid-pair transfers are reachable in the considered time of flights.

\subsection{Solar–sail surrogate}
\label{sec:ml_surrogate_solar_sail}

In the solar–sail case (Sec.~\ref{sec:solar_sailing}), the scalar quantity of interest is the minimum sail–strength parameter \(k_0^*\) that renders a transfer feasible within a fixed time of flight.
For each sampled transfer we solve the PMP boundary–value problem with parameter stationarity, obtaining an optimal value \(k_0^*\) and the associated state–costate history.
The value of \(k_0^*\) directly encodes reachability in terms of sail mass or sail area: smaller values correspond to larger admissible spacecraft masses for a given sail area~\cite{mcinnes2004solar}.

We construct a feature vector \(\bm{\phi}\) encoding the initial and final heliocentric states, the time of flight, and coarse information about the heliocentric distance range covered by the transfer.
The surrogate model is then trained to approximate \(k_0^*\) itself.

\label{sec:ml_architectures}

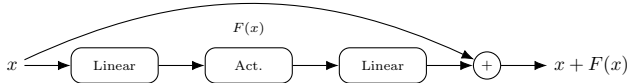
\begin{figure}[t]
\centering
\begin{tikzpicture}[
    scale=0.72,
    every node/.style={transform shape},
    block/.style={
        draw,
        rounded corners,
        minimum width=1.6cm,
        minimum height=0.65cm,
        align=center,
        font=\scriptsize
    },
    sum/.style={
        draw,
        circle,
        minimum size=0.5cm,
        inner sep=0pt,
        font=\scriptsize
    },
    node distance=0.7cm and 0.85cm,
    >=Latex
]

\node (x) {$x$};

\node[block, right=of x] (c1) {Linear};
\node[block, right=of c1] (r1) {Act.};
\node[block, right=of r1] (c2) {Linear};

\node[sum, right=of c2] (add) {$+$};

\node[right=of add] (y) {$x + F(x)$};

\draw[->] (x) -- (c1);
\draw[->] (c1) -- (r1);
\draw[->] (r1) -- (c2);
\draw[->] (c2) -- (add);
\draw[->] (add) -- (y);

\draw[->] (x) to[bend left=25] (add);

\node[above=0.1cm of r1] {\scriptsize $F(x)$};

\end{tikzpicture}
\caption{Residual block}
\label{fig:residual-architecture}
\end{figure}

We consider four classes of fully connected neural networks mapping \(\bm{\phi}\) to a logit target:

\paragraph{Residual MLP (ResNet).}
The main architecture used in this work is a residual multilayer perceptron, whose residual blocks are shown in Fig.~\ref{fig:residual-architecture}.

An input projection maps \(\bm{\phi}\) to a hidden representation of width \(d_h\), followed by a stack of residual blocks of the form
\begin{equation}
    \bm{h}_{\ell+1}
    =
     W_{2,\ell}\,\sigma(W_{1,\ell}\bm{h}_\ell + \bm{b}_{1,\ell})
    + \bm{b}_{2,\ell} + \bm{h}_\ell ,
\end{equation}
where \(\sigma\) is a nonlinearity (we use \texttt{SELU}, \texttt{SiLU}, \texttt{Softplus}, or \texttt{GELU}) and \(\ell\) indexes the residual blocks.
A final linear layer maps the last hidden state to the scalar output.

\paragraph{Plain MLP.}
As a strong non–residual baseline we employ a conventional MLP with \(L\) hidden layers of width \(d_h\), using the \texttt{SELU} and \texttt{ReLU} activations in all hidden layers and Kaiming–normal initialisation.

\paragraph{SIREN network.}
Finally, we include a sinusoidal representation network (SIREN)~\cite{sitzmann2020implicit}, consisting of a first SineLayer that maps \(\bm{\phi}\) to a hidden width \(d_h\), followed by several hidden SineLayers with the same width and a final linear output layer.
Each SineLayer applies a linear transformation followed by a scaled sinusoidal activation ($\sin(\omega_0 \cdot$)), with ($\omega_0$) controlling the effective frequency and the weight initialisation range.
We use the initialisation rules proposed in~\cite{sitzmann2020implicit} and treat \(\omega_0\) as a hyperparameter.

\subsection{Training objective and optimisation}
\label{sec:training_objective}

For the two–body reachability surrogate we adopt a binary–classification objective.
Let \(f_\theta(\bm{\phi})\in\mathbb{R}\) denote the logit output of a network with parameters \(\theta\).
The loss function is the binary cross–entropy with logits,
\begin{equation}
    \mathcal{L}(\theta)
    =
    \frac{1}{N}\sum_{i=1}^N
    \mathrm{BCEWithLogits}\bigl( f_\theta(\bm{\phi}_i), y_i\bigr),
\end{equation}
with a small amount of label smoothing (here \(5\%\)) applied to both classes to mitigate overconfidence and reduce the impact of mislabeled samples.
During validation and testing, we threshold the sigmoid of the logit at \(0.5\) to obtain hard predictions and report accuracy, precision, recall, and F\(_1\)–score, together with the confusion–matrix entries.

In all experiments we use the AdamW optimiser with weight decay, an initial learning rate of order \(10^{-3}\), and a cosine–annealing schedule down to a small floor value.
Mini–batch training is employed, with batch sizes in the few–thousand range.
The best checkpoint is selected based on the lowest validation loss, and all reported test metrics correspond to that checkpoint.

For the solar–sail surrogate, the same optimization setup is used as the low-thrust electric propulsion scenario.
In this case the network output approximates the scalar field $k_0^*(\bm{\phi})$, which plays the same role for solar sails as $m_{\textrm{MIM}}$ does for electric low-thrust transfers.

\subsection{Why residual networks outperform SIREN and plain MLPs}
\label{sec:resnet_advantages}

Across both the two–body reachability and the solar–sail regression experiments, residual MLPs consistently achieve lower validation loss, higher classification F\(_1\)–scores, and tighter error distributions than SIREN and plain MLP architectures of comparable width and depth.\,

This empirical observation is consistent with several known advantages of residual architectures.

First, residual networks are well–suited to such functions because each block learns a perturbation of the identity, so the layerwise Jacobian takes the form
\begin{equation}
    J_{\ell+1}
    =
    \frac{\partial \bm{h}_{\ell+1}}{\partial \bm{h}_\ell}
    \approx I + J_{F_\ell},
\end{equation}
with \(F_\ell\) the residual mapping.
The resulting product of Jacobians across layers tends to remain better conditioned, keeping singular values closer to one and alleviating the vanishing– and exploding–gradient problems that affect deep plain networks~\cite{he2016deep}.
This leads to more stable optimisation and smoother decision boundaries, which also translate into improved generalisation in regions with sparse training data.

Second, the residual parameterisation naturally biases the network towards learning corrections around a simple baseline mapping.
A residual MLP can be interpreted as the integration of small vector-field updates, which yields stable forward dynamics and well-conditioned backward sensitivity.

Third, while SIREN networks are highly expressive and excel at representing high–frequency structure, their sinusoidal activations tend to produce highly oscillatory Jacobians, which can sometimes lead to unstable training results and/or poorer generalization capabilities.

Plain MLPs instead would either require substantially deeper architectures, which exacerbates optimisation difficulties, or significantly wider layers to reach comparable accuracy, increasing computational cost.

Finally, from a practical standpoint, the residual architectures show markedly more robust behaviour under changes in hyperparameters and random seeds.
A larger fraction of training runs converge to high–performing solutions without manual tuning, and the spread in validation metrics across independent runs is reduced.

\section{Experiments}

\subsection{Two–body electric low–thrust reachability}
\begin{table}[t]
\centering
\resizebox{\columnwidth}{!}{%
\begin{tabular}{llrrrrrrrrr}
\toprule
family & activation & N & accuracy & precision & recall & F1 & loss\\
\midrule
mlp & softplus & 64 &  0.8493 & 0.8397 & 0.8537 & 0.8466 & 0.3461 \\
residual & gelu & 64 &  0.9773 & 0.9750 & 0.9785 & 0.9768 & 0.0808  \\
residual & relu & 64 & 0.9730 & 0.9703 & 0.9744 & 0.9724 & 0.0914 \\
residual & selu & 64 &  0.9730 & 0.9703 & 0.9744 & 0.9724 & 0.0914\\
\textbf{residual} & \textbf{softplus} & \textbf{64} & \textbf{0.9777} & \textbf{0.9749} & \textbf{0.9795} & \textbf{0.9772} & \textbf{0.0815} \\
residual & softplus & 128 &  0.9791 & 0.9769 & 0.9802 & 0.9786 & 0.0774\\
residual & softplus & 256 &  0.9791 & 0.9769 & 0.9802 & 0.9786 & 0.0774  \\
\textbf{residual} & \textbf{softplus} & \textbf{512} &  \textbf{0.9785} & \textbf{0.9763} & \textbf{0.9796} & \textbf{0.9780} & \textbf{0.0789} \\
siren & gelu & 64 & 0.9290 & 0.9248 & 0.9300 & 0.9274 & 0.1881 \\
siren & silu & 64 & 0.9301 & 0.9254 & 0.9316 & 0.9285 & 0.1856 \\
siren & softplus & 64 &  0.9290 & 0.9248 & 0.9300 & 0.9274 & 0.1881 \\
\bottomrule
\end{tabular}%
}
\caption{Validation metrics for different architectures and activations for the electric low-thrust propulsion case.}
\label{tab:lt_architectures}
\end{table}
These results were generated leveraging a two-body problem dynamics where the central body is the Sun, with a maximum thrust of 0.1 N, and specific impulse $I_{\textrm{sp}}\in [1000,5000]$ s. In terms of starting and target orbits, these were generated within the following Keplerian elements: semi-major axis $a\in[0.5,3.5]$ AU, eccentricity $e\in [0,1)$, inclination $i\in[0,2 \pi]$, true anomaly $\nu\in [0,2\pi]$, right ascension of the ascending node $\Omega \in [0,2\pi]$, and argument of perigee $\omega \in [0,2\pi]$. 0, 1, 2, and 3 revolutions transfers are generated, with equal proportions in the training, validation, and test sets. A transfer is deemed feasible if the maximum initial mass is above the spacecraft mass (here assumed to be $1500$ kg). Once a solution is generated, the same homotopy strategy as detailed in \cite{zhong2026} is used to continue the family of solutions.
Tab.~\ref{tab:lt_architectures} reports validation metrics for different architectures and activation functions on the electric low–thrust reachability classification task.
Among models with the smallest hidden width ($N=64$), the residual network with \texttt{Softplus} activation achieves the best validation performance, with accuracy $0.9777$, precision $0.9749$, recall $0.9795$, F$_1$–score $0.9772$, and cross–entropy loss $0.0815$.
This configuration is highlighted in bold in the table and already provides a very competitive trade-off between accuracy and number of learnable parameters.

Increasing the width of the residual \texttt{Softplus} network further improves the metrics, but with diminishing returns.
The best overall model in terms of validation loss is the residual \texttt{Softplus} network with $N=512$ neurons per hidden layer, which attains accuracy $0.9785$, precision $0.9763$, recall $0.9796$, F$_1$–score $0.9780$, and loss $0.0789$.
While this represents a slight improvement over the $N=64$ counterpart, the gains are not dramatic when compared to the substantial increase in parameter count and computational cost, suggesting that relatively compact residual networks already capture most of the structure of the dual reachability indicator.

On the held–out test set (comprising $10\%$ of the overall data), the selected residual \texttt{Softplus} model yields the following confusion matrix:
\begin{equation}
\mathrm{CM}_\text{test} =
\begin{bmatrix}
468418 & 11138\\
9143 & 445759
\end{bmatrix},
\end{equation}
where the rows correspond to the true labels (unreachable, reachable) and the columns to the predicted labels.
This corresponds to an overall test accuracy of approximately $0.978$, precision of about $0.976$, recall of about $0.980$, and F$_1$–score of about $0.978$, in line with the validation metrics.
Most errors occur near the reachability boundary: the number of false positives (predicted reachable but actually infeasible) and false negatives (predicted infeasible but actually reachable) is small compared to the number of correctly classified transfers, and their distribution reflects the intrinsic ambiguity of borderline cases.

\begin{figure}[tb]
  \centering
    \includegraphics[width=0.5\textwidth]{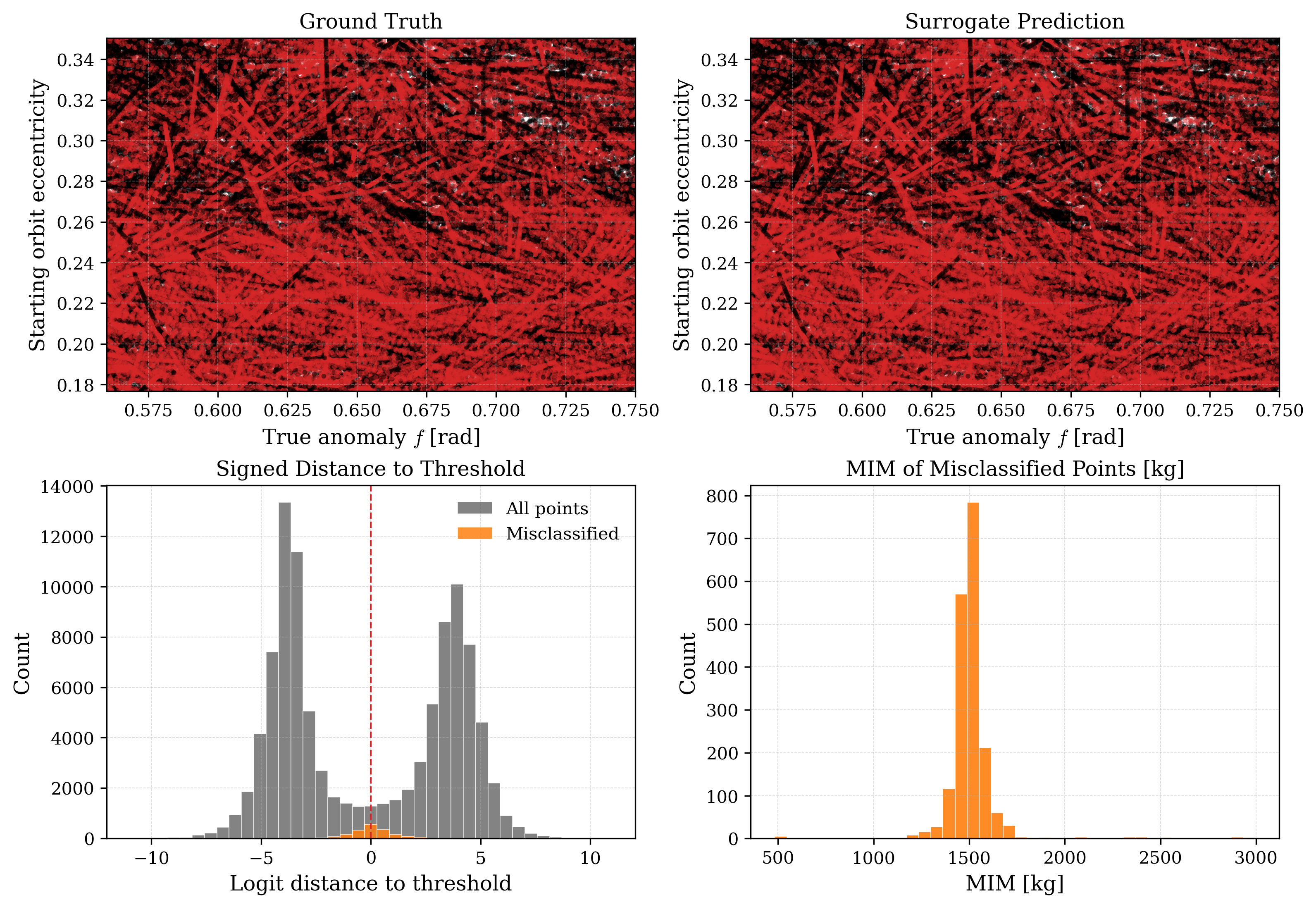}
\caption{Comparison of ground truth and surrogate reachability on the test set. The surrogate captures the main feasible/infeasible structure and sharp decision boundary, with most errors and low-confidence predictions occurring near the threshold, particularly close to the mass limit $m_{\mathrm{sc}}=1500~\mathrm{kg}$.}
\label{fig:lt_reachability_qualitative}
\end{figure}

To illustrate this behaviour qualitatively, Fig.~\ref{fig:lt_reachability_qualitative} shows a 2×2 panel restricted to a subset of the test set.
The top–left panel displays the ground–truth reachability labels in the plane of departure true anomaly and initial orbit eccentricity, while the top–right panel shows the corresponding surrogate predictions for the same samples.
The surrogate reproduces the main structure of the reachable and unreachable regions, including sharp transitions across the boundary, with visually small regions of disagreement.

The bottom–left panel depicts the signed distance of the logits to the decision threshold, i.e.\ the signed margin with respect to the $0.5$ probability level.
The distribution of these margins is clearly bimodal: confidently classified reachable and unreachable transfers populate two well-separated lobes, while misclassified cases concentrate near the decision boundary where the margin is small in absolute value.
This is consistent with the idea that most classification errors occur for transfers that are intrinsically close to the feasibility threshold.

Finally, the bottom–right panel shows the distribution of misclassified test samples in terms of the true maximum initial mass.
As expected, these points cluster around the threshold $m_{\mathrm{sc}}=1500~\mathrm{kg}$: transfers that are only marginally feasible or infeasible are the ones that are most difficult to classify, both for the surrogate.
This behaviour is desirable from a mission-design standpoint, as it indicates that the surrogate rarely misclassifies clearly reachable or clearly unreachable targets, and concentrates its uncertainty in a narrow band around the operational mass limit.

\subsection{Solar–sail surrogate performance}
In the solar-sail case, the GTOC13 problem\footnote{\url{https://gtoc.jpl.net/gtoc13/the-problem/}} is used as the simulation environment (“gym”), with Altaira as the central body. The orbital parameters follow the same bounds as in the original formulation, except for the semi-major axis, which is extended to the range 0.01–8 AU. 
The sail area and spacecraft mass used to compute the threshold sail strength, and thereby determine target reachability, are 15,000 m$^2$ and 500 kg, respectively.
For these experiments, we reuse the same residual \texttt{Softplus} architecture identified as best in the electric low–thrust study.
On the validation set, the best configuration attains loss $0.0364$, precision $0.9922$, accuracy $0.9945$, and recall $0.9927$.
From these values we compute an F$_1$–score of 0.9924, indicating a very high-quality surrogate that almost perfectly separates feasible from infeasible transfers when thresholded at the appropriate $k_0$ level.

Taken together with the electric low–thrust results, this confirms that residual networks with smooth activations provide a robust architecture choice across both reachability problems studied here.
Once trained, the solar–sail surrogate can be queried at negligible cost to evaluate approximate reachability over large sets of target states, effectively turning the MIM-based dual formulation into a practical, data-driven reachability oracle for preliminary sail-mission design.

\section{Conclusions}

This work introduced a dual formulation of low-thrust reachability based on the notion of maximum initial mass (MIM) for electric propulsion and an analogous minimum sail-strength parameter for solar-sail transfers. By reformulating the classical reachability question into the evaluation of the maximum initial mass scalar field, the proposed approach replaces the construction of a high-dimensional reachable set with the computation of a well-defined, continuous, and smoother performance indicator. This reformulation retains the full feasibility information of the original problem while significantly improving its numerical structure, particularly near the boundary of the reachable set where the corresponding dual minimum time problem tends to exhibit less stability and convergence issues.

We derived indirect optimal-control formulations for both electric low-thrust and solar-sail dynamics using Pontryagin’s Minimum Principles. In the electric case, the resulting boundary-value problem is formulated via a forward–backward shooting scheme. For solar sailing, we wrote the PMP to include parameter stationarity conditions, enabling the consistent treatment of the sail-strength parameter within the same framework. These developments provide a unified viewpoint in which reachability is encoded through scalar optimality conditions.

Building on these dual formulations, we investigated data-driven surrogate models that approximate the resulting scalar reachability indicators. Across both problem classes, residual multilayer perceptrons with smooth activations, particularly Softplus-based architectures, consistently outperformed plain MLPs and SIREN networks in terms of accuracy and training stability. The learned surrogates accurately reproduce the structure of the reachable set, making them well-suited as practical reachability oracles.

Overall, the combination of indirect MIM formulations and neural surrogates provides an efficient and scalable framework for preliminary mission design under low-thrust dynamics. The resulting approach enables rapid feasibility assessment without repeated optimal-control solves, substantially reducing computational cost while preserving high accuracy. 
\bibliographystyle{ISSFD_v01}
\bibliography{bibliography}

\end{document}